\documentclass[preprint,12pt]{elsarticle}

\usepackage{amssymb}
\usepackage{url}
\usepackage{color}
\usepackage{ulem}
\usepackage{listings}

\begin{document}

\begin{frontmatter}



\title{Information transmission on hybrid networks}

\author[label1]{Rongbin Chen}

\cortext[cor1]{200 Xiaolingwei, Nanjing 210094, China. Tel: +8613915966537.}
\author[label2]{Wei Cui}
\author[label1,label4]{Cunlai Pu\corref{cor1}}
\ead{pucunlai@njust.edu.cn}
\author[label3]{Jie Li}
\author[label1]{Bo Ji}
\author[label4]{Konstantinos Gakis}
\author[label4]{Panos M. Pardalos}

\address[label1]{Department of Computer Science and Engineering, Nanjing University of Science and Technology, Nanjing 210094, China}
\address[label2]{EMC corporation, Beijing 100027, China}
\address[label3]{Hefei National Laboratory for Physical Science at Microscale, University of Science and Technology of China, HeFei 230026, China}
\address[label4]{ Department of  Industrial and Systems Engineering, University of Florida, Gainesville, FL 32611, USA}

\begin{abstract}
Many real-world communication networks often have hybrid nature with both fixed nodes and moving modes, such as the mobile phone networks mainly composed of fixed base stations and mobile phones. In this paper, we discuss the information transmission process on the hybrid networks with both fixed and mobile nodes. The fixed nodes (base stations) are connected as a spatial lattice on the plane forming the information-carrying backbone, while the mobile nodes (users), which are the sources and destinations of information packets, connect to their current nearest fixed nodes respectively to deliver and receive information packets. We observe the phase transition of traffic load in the hybrid network when the packet generation rate goes from below and then above a critical value, which measures the network capacity of packets delivery. We obtain the optimal speed of moving nodes leading to the maximum network capacity. We further improve the network capacity by  rewiring the fixed nodes and by considering the current load of fixed nodes during packets transmission. Our purpose is to optimize the network capacity of hybrid networks from the perspective of network science, and provide some insights for the construction of future communication infrastructures.
\end{abstract}


\end{frontmatter}


\section{Introduction}

Nowadays, we greatly depend on the communication infrastructures, such as mobile phone networks, Internet, wireless sensor networks, or most recently Internet of Things, to deliver and receive text, voice, and videos, etc. The expansion of these infrastructures and the increasing interconnection between them make it hard for us to understand and control these infrastructures. As a consequence we encountered so many network problems, such as the attacks on the Internet \cite{Lis}, large-scale spreading of computer viruses \cite{tangc,Yanghx16,PUC16,puc15}, network congestion \cite{welzl,jiy},  and many others.

One of the fundamental issues is what's the maximum amount of flow a communication network can carry \cite{liqc}. In the past decade, researchers from the area of network science focused on discussing the delivery capacity of various complex networks based on the methods from statistical physics \cite{schen12}. Network capacity is defined as the critical packets generation rate, when the packets generation rate goes from below this critical value to above this value, the network undergoes a phase transition from free flow to traffic congestion \cite{zhaol,arenas01}. Furthermore, network capacity can be obtained through estimation in terms of node delivery capability, betweenness centrality \cite{brandesu,magaia}, and network size \cite{guimerar}. In addition, network capacity also has relations with other network properties like link density and degree distribution \cite{schen12}.

Much effort has been devoted to finding various strategies to improve the network capacity. One type of these strategies is to optimize network topological structures or network resources.
 For example, Liu et al \cite{zliu} found that by removing some of the links between the core nodes, the network capacity can be effectively increased. Similarly, Zhang et al \cite{gqzhang}  obtained that removing the links between nodes of large betweenness also increases the network capacity. Huang et al \cite{whuang}  proposed an effective strategy which adds shortcut links between nodes that have the largest shortest path lengths. Yang et al \cite{yanghx08} studied the optimal allocation of node delivery capacity in order to achieve the maximum network capacity, and they proposed a degree-based allocation strategy, which is controlled by a free parameter. They further obtained the optimal parameter through both simulation and analysis.

The optimization of network structures often costs much or is unfeasible in real situations. The other type of strategies is to optimize the routing strategies \cite{pucl12,wangk11,pucl13,sheny06,wang026}, which is more economical and applicable to real situations.
Yan et al \cite{gyang046}  proposed an efficient routing strategy in which traffic flow would deliberately avoid passing through the large degree nodes and thus the traffic load is redistributed from large degree nodes to small degree nodes. Danila et al \cite{danilab}  provided a heuristic algorithm to reduce the maximum betweenness of nodes to improve network capacity. Ling et al \cite{ling2010}  proposed a global dynamic routing strategy to select the optimal path, in which the sum of the node queue lengths  is the smallest.
Wang et al \cite{wangwx016}  proposed a local dynamic routing strategy, which not only considers the degree of nodes but also  the load of neighboring nodes.

Most recently, some attention has been shifted to  multilayer networks \cite{Boccaletti14,puc1626}.
Zhou et al \cite{zhouj13} studied the optimal routing on multilayered networks composed of a wireless network and a wired network. They developed a recurrent algorithm which is better than the shortest path algorithm in terms of transport capacity. Du et al \cite{du72}  considered the multilayered network, where the lower layer is a lattice and the upper layer is a scale-free network. They assigned different transmission costs to each of the two layers and investigated the optimal coupling of the two network layers  to obtain the maximum network capacity.
Tan et al \cite{tan14} investigated how the interconnection of the BA scale-free networks as well as the interconnection of the autonomous system (AS) graphs of the Internet in South Korea and Japan affect the traffic flow, and found that assortative coupling is less susceptible to traffic congestion than random coupling and disassortative coupling when the node processing capacity is allocated based on node usage probability.

So far, when discussing communication networks, we usually assume that all  nodes in the networks are the same \cite{schen12}. For example, every node is both router and computer, and all nodes are either immobile or mobile. However, these assumptions are not fit for many real-world communication networks. For instance, in mobile communication networks there are different kinds of nodes, and the two main types are base stations and mobile phones \cite{kgomez}. For the based station, it is fixed and plays the role of information forwarding, while for the mobile phone, it is the source and destination of information, and  has a certain level of mobility. Thus it is necessary to discuss the information transmission process on such hybrid communication networks from the perspective of network science.

In this paper, we model the hybrid communication network consisting of two kinds of nodes: fixed based stations and mobile users. Then we simulate the information transmission process on this hybrid network. We focus on the network capacity and how it is affected by the factors such as user moving speed and number of users. Moreover, we try to improve the network capacity by optimizing the network topological structure, as well as by considering the node load in the design of routing protocols.

\section{The network model}
The hybrid network is embedded  in a 2-dimensional plane, which consists of an information-carrying backbone and many users. The topological structure of the information-carrying backbone is a regular lattice. The length of each link of the lattice is 1. The lattice forms a bounded square area for the users' movement. The code for generating the lattice is given in \textbf {Appendix A}. We allocate a base station  at each site of the lattice. Thus, each base station has four neighbors with one in each of the four directions.   At time $t=0$, $n$ mobile users are randomly distributed  within the square area. The moving speed of all the users is a constant value $v$ during the whole time. The moving direction of each user is randomly generated within range [0, 2$\pi$] at the beginning, and the moving direction will not change until the user run into the boundaries of the lattice, when the moving direction is regenerated. At each time step, the users only connect the nearest base stations. Because of the users' movement, the links between users and base stations change with time. The  topological structure of the hybrid network is shown in Fig. 1(a).

\subsection{Network optimization}
In the above, the backbone for information transmission is a lattice. However, we know that the topological structures of many real communication networks are neither regular nor random. Instead, they have small-world and scale-free properties \cite{bal09,bal16}. Enlightened by this, we perturb the lattice structure by rewiring some of its links to improve the topological randomness of the backbone. Here we provide a rewiring strategy, namely direction-based rewiring strategy (DBRS), to optimize the network structure. In the DBRS, every base station  rewires its right link and down link respectively with probability $P$,  the new right (down) base station of the right (down) link is randomly chosen from the right (down) direction, shown in Fig. 1(b).  The code of DBRS is given in \textbf{Appendix B}.  Note that we can also choose to rewire the links in the other directions. For example,  we can also rewire the left and up links, then the new base stations of the links are chosen from the left and up directions respectively. In addition, we do not select new  base stations among the whole lattice, since we want to make sure the backbone is still a connected component after perturbation.

\begin{figure}
 \centering
\includegraphics[width=4in,height=1.8in]{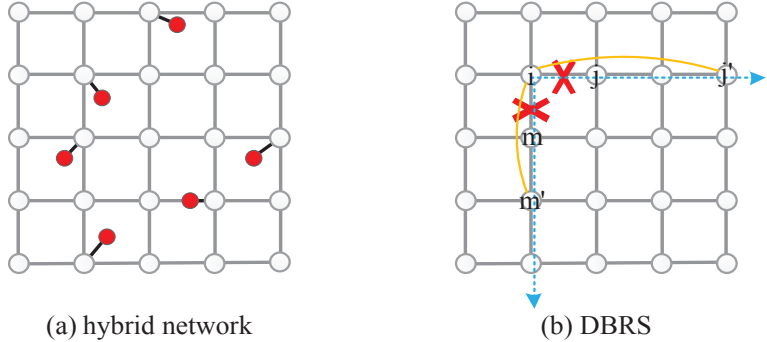}
\caption{Examples of (a) the hybrid network and (b) the DBRS for illustration purpose.  The hybrid network  is composed of a backbone (grey lattice) and mobile users (red nodes). The backbone consists of connected base stations with one base station on each site of the lattice.  The length of each link of the lattice is 1. The moving users will connect  the current nearest base stations (as the gateways) during the packets delivery. The links between base stations are fixed, while the links between users and base stations are dynamical.  In the DBRS, every node will randomly rewire the right link (in the right direction)  and down link (in the down direction)  with probability $P$ respectively.  For instance, the right link of node $ i$ is originally $\langle i,j \rangle$, after random  rewiring, the right link becomes $\langle i,j' \rangle$. The down link of node $i$ is initially $\langle i, m\rangle$, after random rewiring, the down link becomes $\langle i, m' \rangle$.
}
\end{figure}

\section{The transmission model}
At each time step, every user generates a packet with probability $\rho$, and thus there are on average $n\rho$ packets  generated in the network, where $ n$ is the total number of users.  The destinations of these new packets are  randomly selected from all the users except the source users. Then, these packets are sent to the nearest base stations (gateways) of the sources. Thereafter, the packets are transmitted in the backbone hop by hop to their destinations. Once the packets arrive at the gateways of their destinations, they will be sent to the destinations directly and removed from the hybrid network. Each base station delivers  at most $C$ packets each time, and it has an infinite queue for buffering the remaining packets, which obeys the first-in-first-out (FIFO) rule. In the backbone, the base stations perform the shortest path protocol to transmit the packets, which means when a base station delivers a packet, the base station will check the routing table, and then send the packet to the next hop on the shortest path to the destination. If there are many shortest paths to the destination, we randomly choose one of them as the transmission path. The shortest path is intuitively the fast path to the destination. However, the shortest paths between different sources and destinations usually intersect at a few large degree nodes, causing the traffic congestion problem and thus decreasing the transmission efficiency.

In order to analyze the transition from free flow to traffic congestion, we use the order parameter presented in previous studies \cite{arenas01}:
\begin{eqnarray}
\eta(\rho)=\lim_{t\rightarrow\infty}\frac{C}{n\rho}\frac{\langle\Delta W\rangle}{\Delta t},
\end{eqnarray}
where $W(t)$ is defined as the number of packets in the network at time $t$, and $\Delta W=W(t+\Delta t)-W(t)$, $\langle...\rangle$ indicating averaging over time window of width $\Delta t$. The order parameter $\eta$ represents the ratio between the increased load and the inserted load after a long time period. Obviously, when $\rho$ is relatively small, the generated packets are balanced with the removed packets, and $\eta$ is close to 0, which means the network is under free flow sate. There is a critical packet generation rate $\rho_c$, when $\rho>\rho_c$, the packets are continuously accumulated in the network, and $\eta$ is greater than zero, which means the network is under congestion state. The network capacity is thus  measured by the critical packet generation rate $\rho_c$.   $\rho_c$ can be obtained  based on the bisection method, the code of which is given in \textbf {Appendix C}.
\subsection{Routing optimization}
To alleviate the traffic congestion and improve the network capacity, we further consider the load of base stations during the selection of transmission paths. Specifically, when delivering a packet, the base station will check the traffic load of next hops in all the shortest paths to the destination. Then, the base station of the smallest traffic load will be chosen as the optimal next hop to deliver the packet.
Let us assume that node $s$ wants to send a packet whose destination is node $t$. Then the optimal next hop from node $s$  to node  $t$ is as follows:
 \begin{eqnarray}
 Next(s,t)=\{p|L(p)=\min (L(p^\ast)),\nonumber \\
 p^\ast \in\{q|H(q,t)+1=H(s,t), q \in g(s)\}\},
\end{eqnarray}
where $g(s)$ is the  neighbor set of node $s$, $L(p)$ is the load of node $p$, and $H(s,t)$ is the shortest path length between node $s$ and $t$.
For comparison purpose, we also consider the base station of the largest traffic load as the next hop in the experiments.

\section{Results}
In this section, we study the phase transition of the traffic flow, measure the network capacity, and discuss how the network capacity are affected by the user moving speed and the number of users, and how we can improve the network capacity through network structural optimization and routing optimization. The number of base stations in the backbone is $N=1024$, the delivery capability of each base station is $C=10$, and these two parameters are fixed in all the experiments. In addition, the shortest path protocol is utilized to route  packets in the experiments.

Firstly, we study the relationship between the packet generation rate $\rho$ and the order parameter $\eta$, as well as the average packet arrival time $T$. The topological structure of the backbone is a lattice.
 In the simulation, we set number of users $n=1000$ and the users' moving speed  $v=0.3$. The results are given in Fig. 2, which are the average of 1000 independent runs. In Fig. 2(a), we see that with the increase of $\rho$, $\eta$ is close to zero at the first stage, and then increases abruptly when $\rho$ surpasses the critical value $\rho_c$ (=0.218) and finally converges.  A clear phase transition of traffic flow from free flow to traffic congestion is observed.
According to the previous results \cite{guimerar}, the critical value of the packet generation rate $\rho_c $ is in general estimated as follows:
\begin{eqnarray}
\rho_c  = \frac{(N-1)C}{B^{*}},
\end{eqnarray}
where $B^{*}$ is the largest node betweenness.  Eq. (3) can not be directly applied to our model, since our model is more complex.  Basicly, $\rho_c$ is also related to $v$ (users are moving) and $n$ (only users generate packets) for our model.  When $n$ is large enough and $v$ is not very large, the base stations will receive a approximately constant rate of new packets from users. In this ideal case, we can assume that the packets are generated by the base stations (not users) with a constant rate. Then, the number of inserted packets for a time step when the traffic is close to congestion can be calculated as follows:
  \begin{eqnarray}
S_c=n\ast\rho_c^n = N\ast\rho_c^N,
\end{eqnarray}
where $\rho_c^n$ is the critical packet generation rate of user, and $\rho_c^N$ is the hypothetical critical packet generation rate of base station. Then, we obtain $\rho_c$ of our model through the following calculation:
\begin{eqnarray}
\rho_c = \rho_c^n = \frac{N(N-1)C}{nB^{*}}.
\end{eqnarray}
The ideal value of $\rho_c$ for settings of the parameters in Fig. 2 is 0.221, which is just a little bit larger than the simulation result of $\rho_c$  (=0.218).
In Fig. 2(b), the shape of the curve is similar to Fig. 2(a). There is also a critical packet generation  number $\rho_c$. When $\rho<\rho_c$, $T$ is around 27, which is the average time  for each packet to reach  the destination.  When $\rho>\rho_c$, $T$ rapidly increases with time and then converges. This indicates that transmission efficiency is greatly affected by the traffic congestion. Note that $\rho_c$ obtained in Fig. 2(a) and (b) are very close. In addition, typical error bars are given  in Fig.  2 to show the deviation of the results.

\begin{figure}
\centering
\includegraphics[width=2.5in,height=2in]{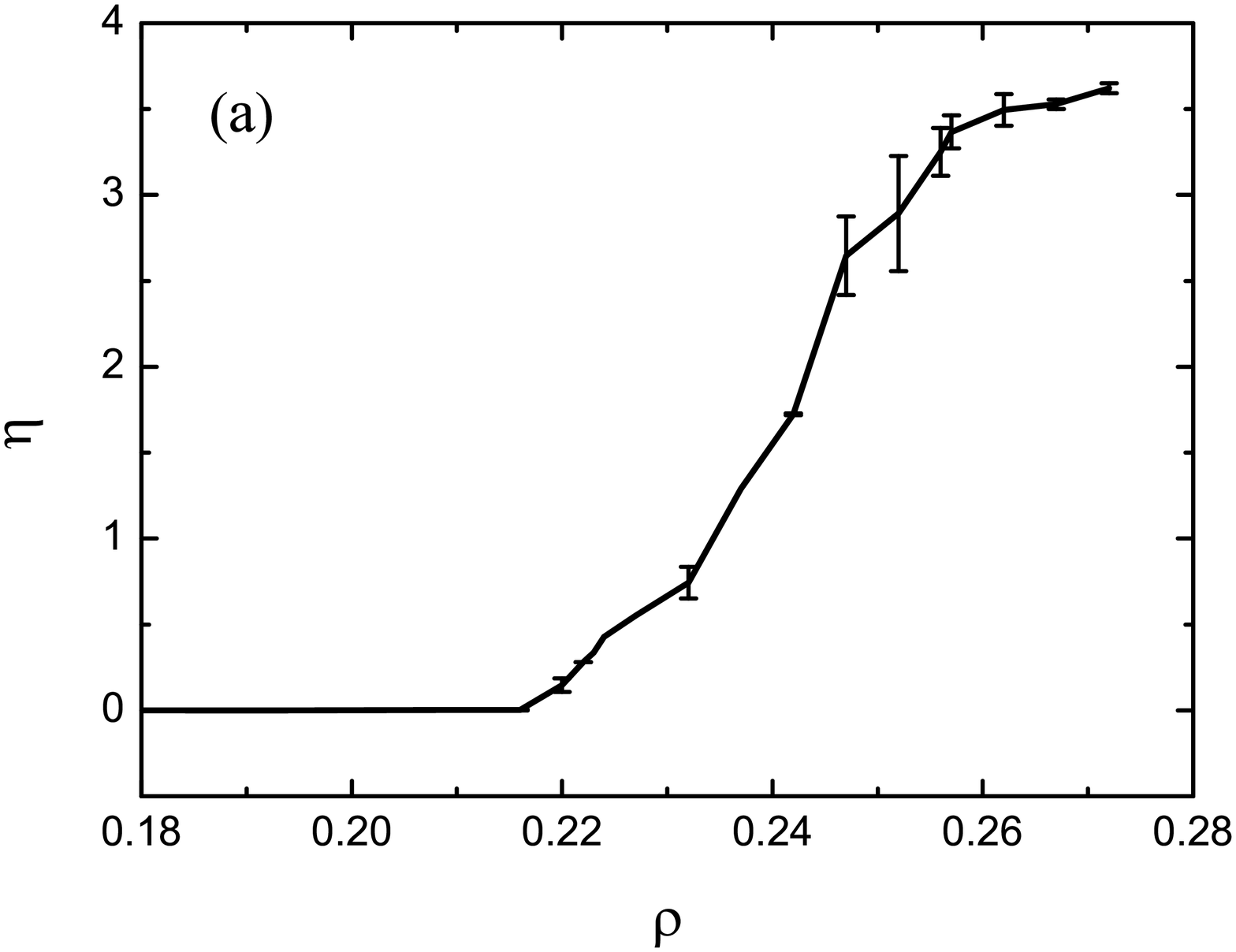}
\includegraphics[width=2.5in,height=2in]{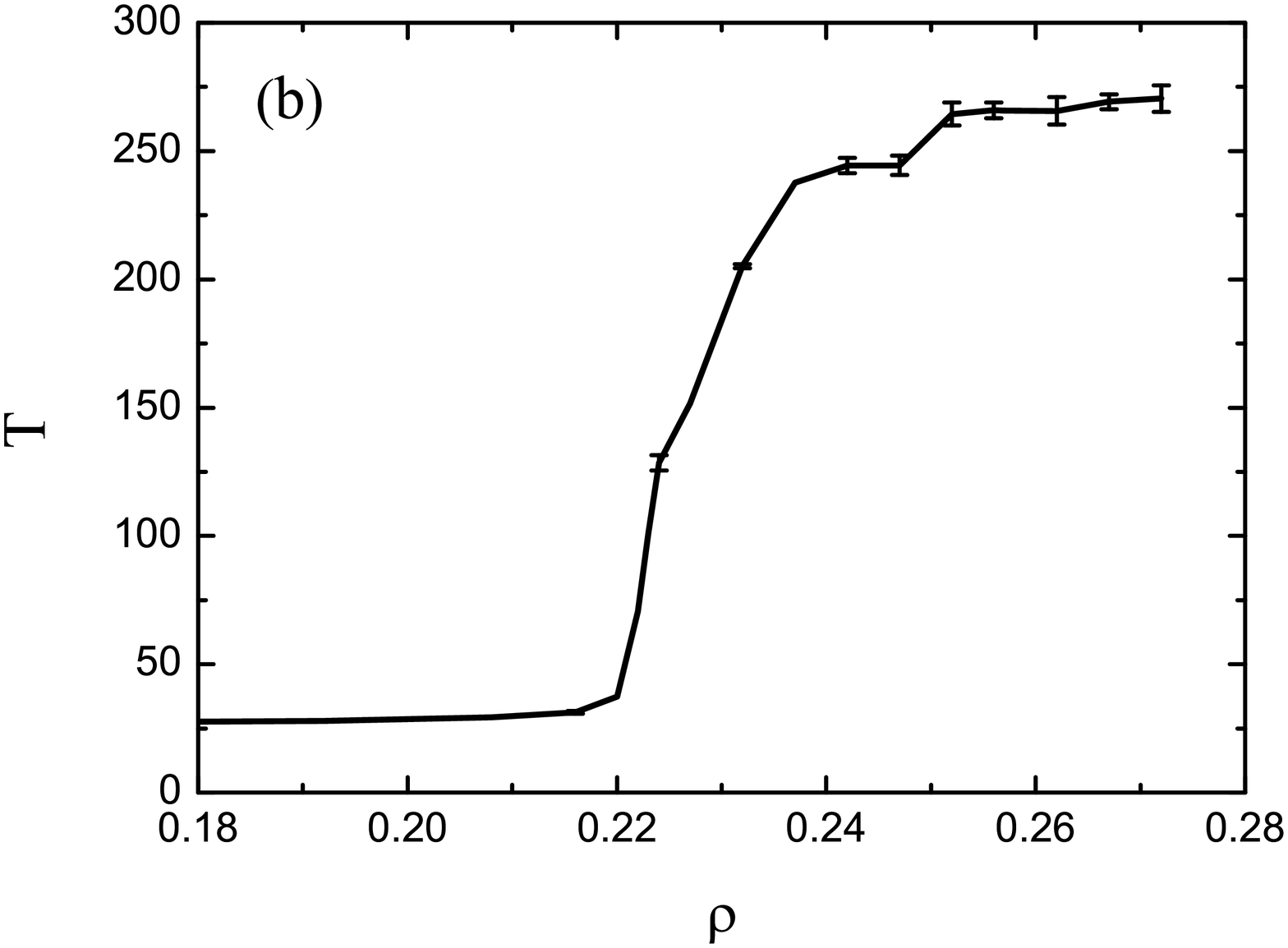}
\caption{(a) Order parameter $\eta$ and (b) average arrival time $T$ vs.  packet generation rate $\rho$. The backbone is a regular lattice. The number of base stations is $ N = 1024$.  The  delivery capacity of base station is $C =10$. The number of mobile users is $n=1000$, and the user moving speed is $v=0.3$. Each data point is the average of 1000 independent runs. }
\end{figure}

Then, we take the critical packet generation rate as the measure of network capacity, and study how to optimize the network structure to improve the network capacity. We test the performance of DBRS which is introduced in the above. In Fig. 3, $ n=100$  and the other parameters are the same as in Fig. 2. We can see that $\rho_c$ slightly decreases first, then increases with $P$, and finally decreases with $P$.  There is an optimal $P$ (=0.25) corresponding to the maximum $\rho_c$ (=2.36). In the DBRS, on one hand, we rewire the links randomly, which introduces the shortcut links between distant base stations. On the other hand,  we reserve some spatial properties of the links, for instance, the right link is also the right link after rewiring. This will make two  physically close base stations also relatively close in the sense of topological distance, and this is a good property for packets transmission since the users' moving direction is generally unchanged.
\begin{figure}
\centering
\includegraphics[width=3in,height=2.3in]{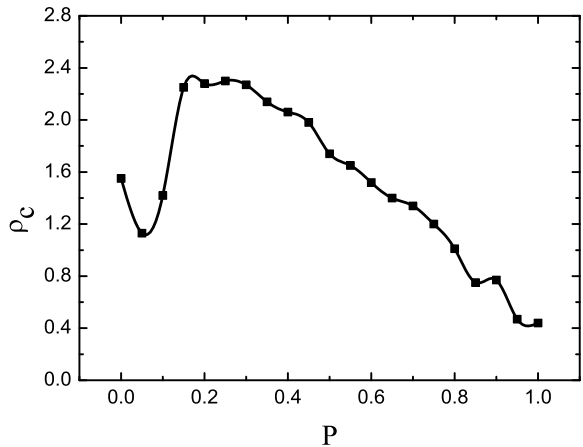}
\caption{Critical packet generation rate $\rho_c$ vs. perturbation probability $P$ for the DBRS.
  The number of base stations is $ N = 1024$.  The delivery capacity of base station is $C =10$. The number of mobile users is $n=100$, and the user moving speed is $v=0.3$. Each data point is the average of 5000 independent runs. }
\end{figure}

 To further explain the results of Fig. 3, we calculate the degree distribution $P(k)$ of base stations and  the load variance $\sigma_L$  of base stations.  Note that the degree  of a base station equals the number of links it has with other base stations (The dynamic links between base stations and users are not considered here). $\sigma_L$ is calculated  as follows:
 \begin{eqnarray}
\sigma_{L}  = \sqrt{\frac{\sum_{i=1}^N (L_{i}-\bar{L})^{2}}{N}},
\end{eqnarray}
 where $L_i$ is the load of base station $i$, and $\bar{L}$ denotes the average load of all the base stations. In the simulation, we set  a relatively small packet generation rate ($\rho=0.3$)  to make the network under  free flow state. The results are given in Fig. 4.
 In Fig. 4(a), we see that with the increase of $P$, the degree distribution $P(k)$ becomes more and more broad. The emergence of large-degree base stations greatly influences the global load distribution of base stations. As shown in Fig. 4(b), the load variance $\sigma_L$ first increases slightly with $P$, then decreases, and finally increases with $P$, the variation trend of which is opposite to that of Fig. 3. This indicates that the load distribution determines the network capacity.
\begin{figure}
\centering
\includegraphics[width=2.5in,height=2in]{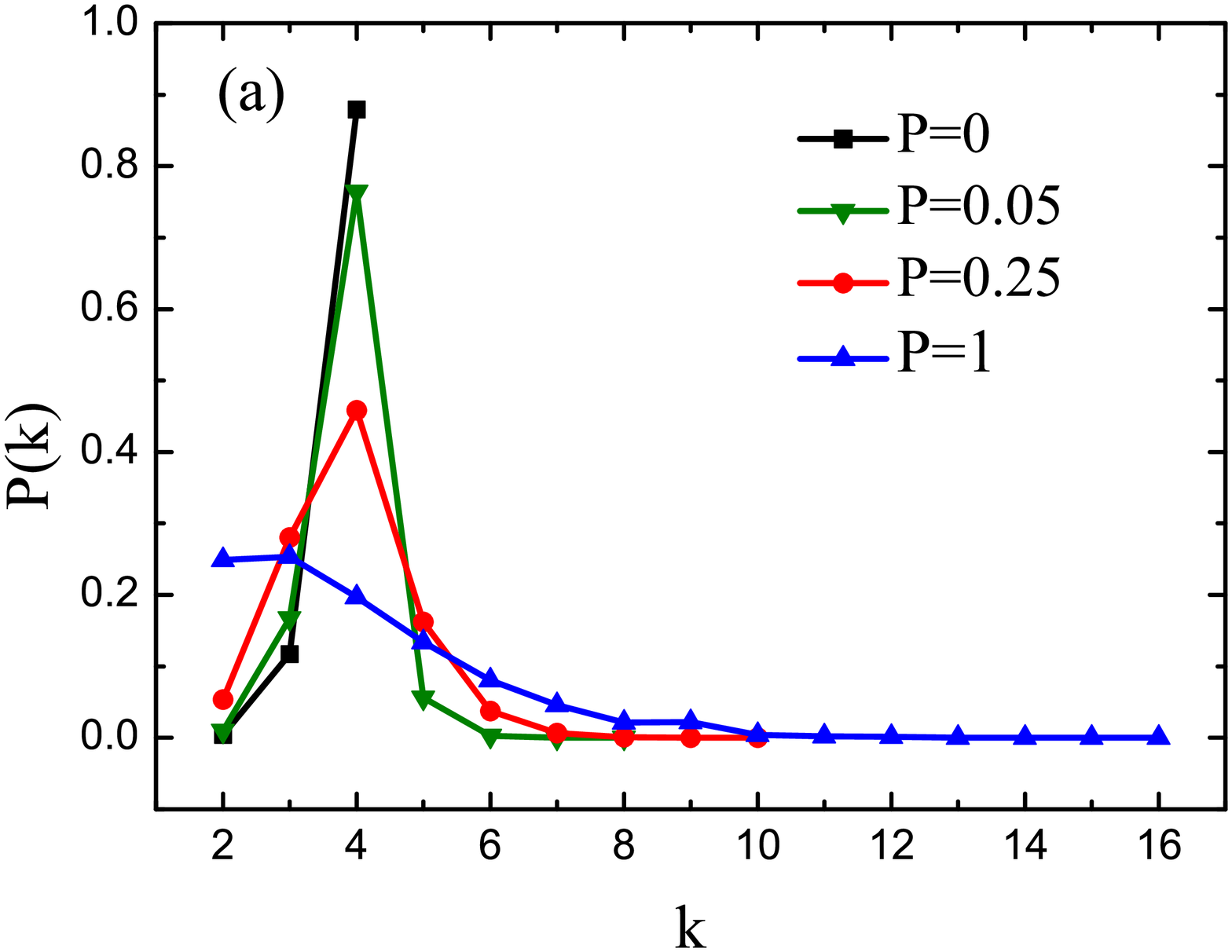}
\includegraphics[width=2.5in,height=2in]{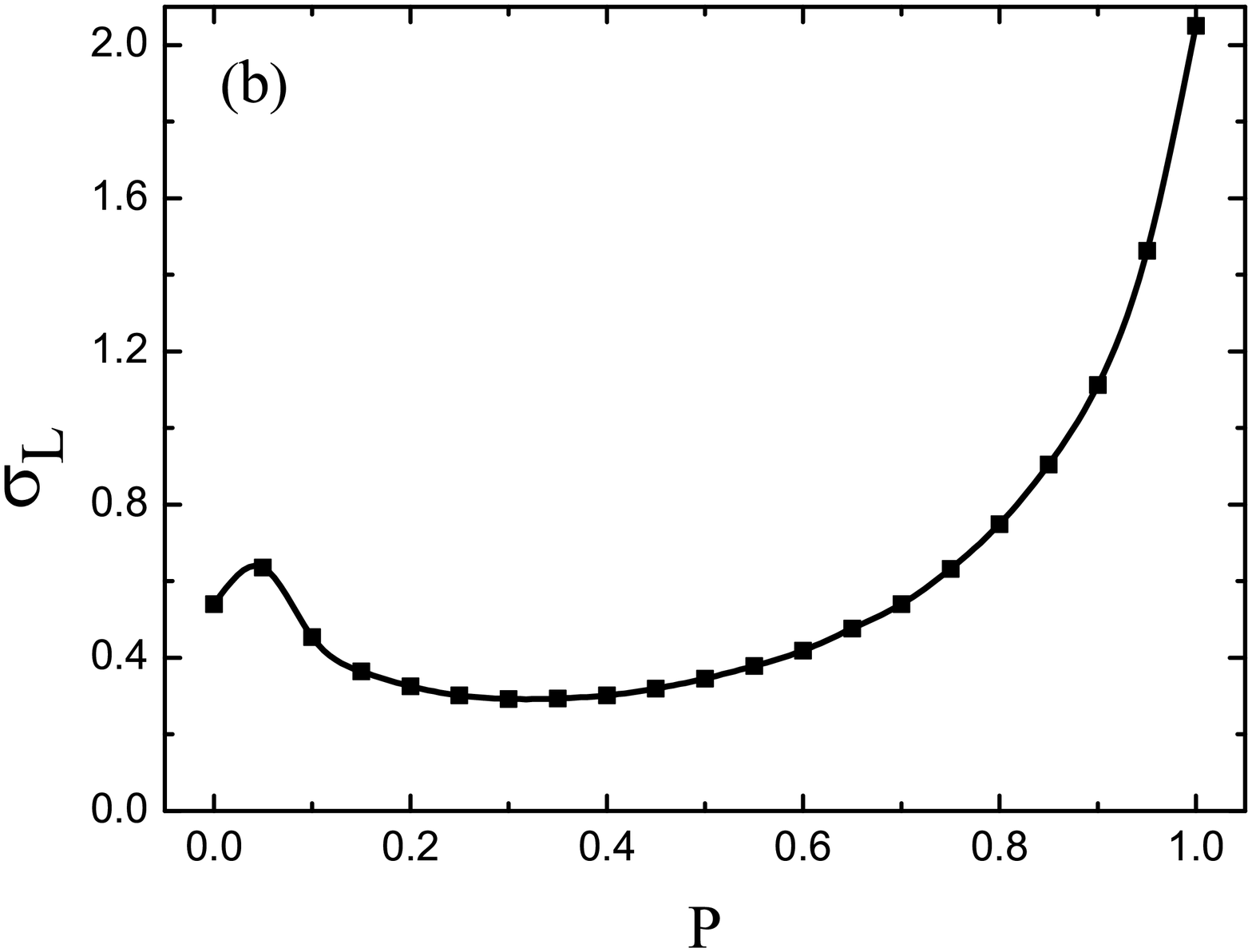}
\caption{(a) Degree distribution $P(k)$ of base stations, and (b) load variance $\sigma_L$ of base stations vs. perturbation probability $P$.  The packet generation rate is $\rho=0.3$, under which there is no traffic congestion. The number of base stations is $ N = 1024$.  The delivery capacity of base station is $C =10$. The number of mobile users is $n=100$, and the user moving speed is $v=0.3$. Each data point is the average of 1000 independent runs. }
\end{figure}

Next, we study the influence of users' moving speed $v$ and number of users $n$ on network capacity. The backbone is optimized by the DBRS with $ P = 0.25$. The results are shown in Fig. 5 and Fig. 6, and each data point is the average of 1000 independent runs. In Fig. 5, we see that $\rho_c$ increases with $v$ first, and then decreases with $v$. When $ v = 0$, all users are immobile. The delivery paths of packets are fixed, and then the packets are prone to accumulate in some large-degree base stations, which are the intersections of many delivery paths. This limits the network capacity.  When $v$ is larger than 0, the users will change connections with base stations during the packets transmission, this makes the packets distributed more fairly among base stations, and thus increases the network capacity. However, when $v$ is too large, the packets are hard to arrive at their destinations, since the destination users change the connections  so often. This results in more time steps to deliver the packets and a decrease of network capacity.

\begin{figure}
\centering
\includegraphics[width=3in,height=2.25in]{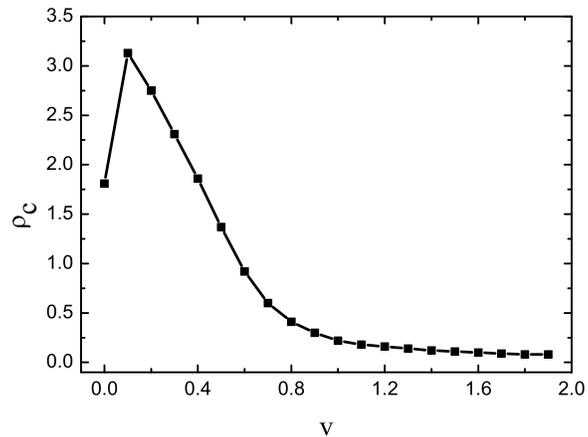}
\caption{Critical packet generation rate $\rho_c$ vs. user moving speed $v$. The backbone is optimized by the DBRS with $ P = 0.25$.
 The number of base stations is $ N = 1024$.   The  delivery capacity of base station is $C =10$. The number of mobile users is $n=100$.   Each data point is the average of 1000 independent runs. }
\end{figure}

\begin{figure}
\centering
\includegraphics[width=2.5in,height=2in]{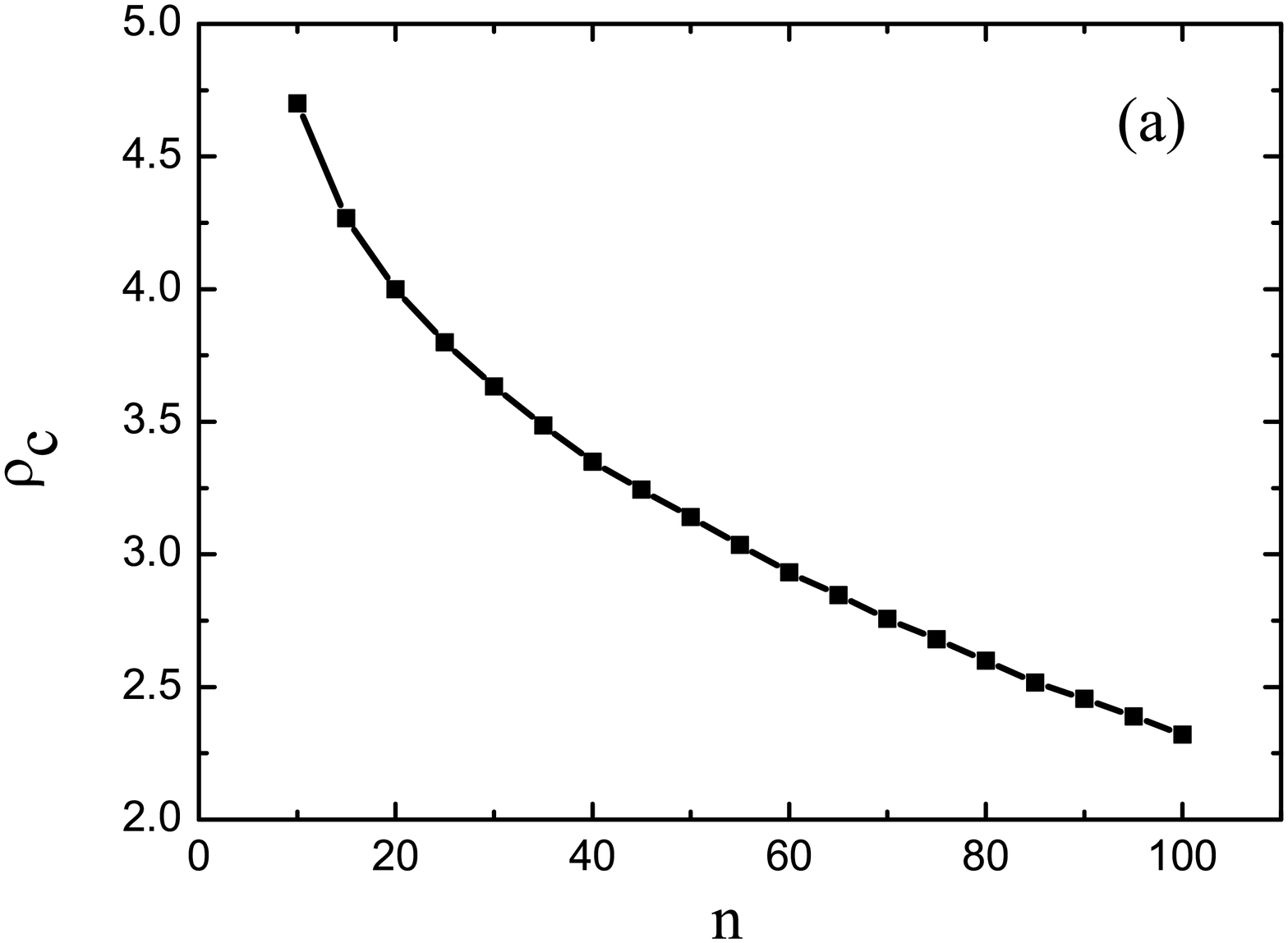}
\includegraphics[width=2.5in,height=2in]{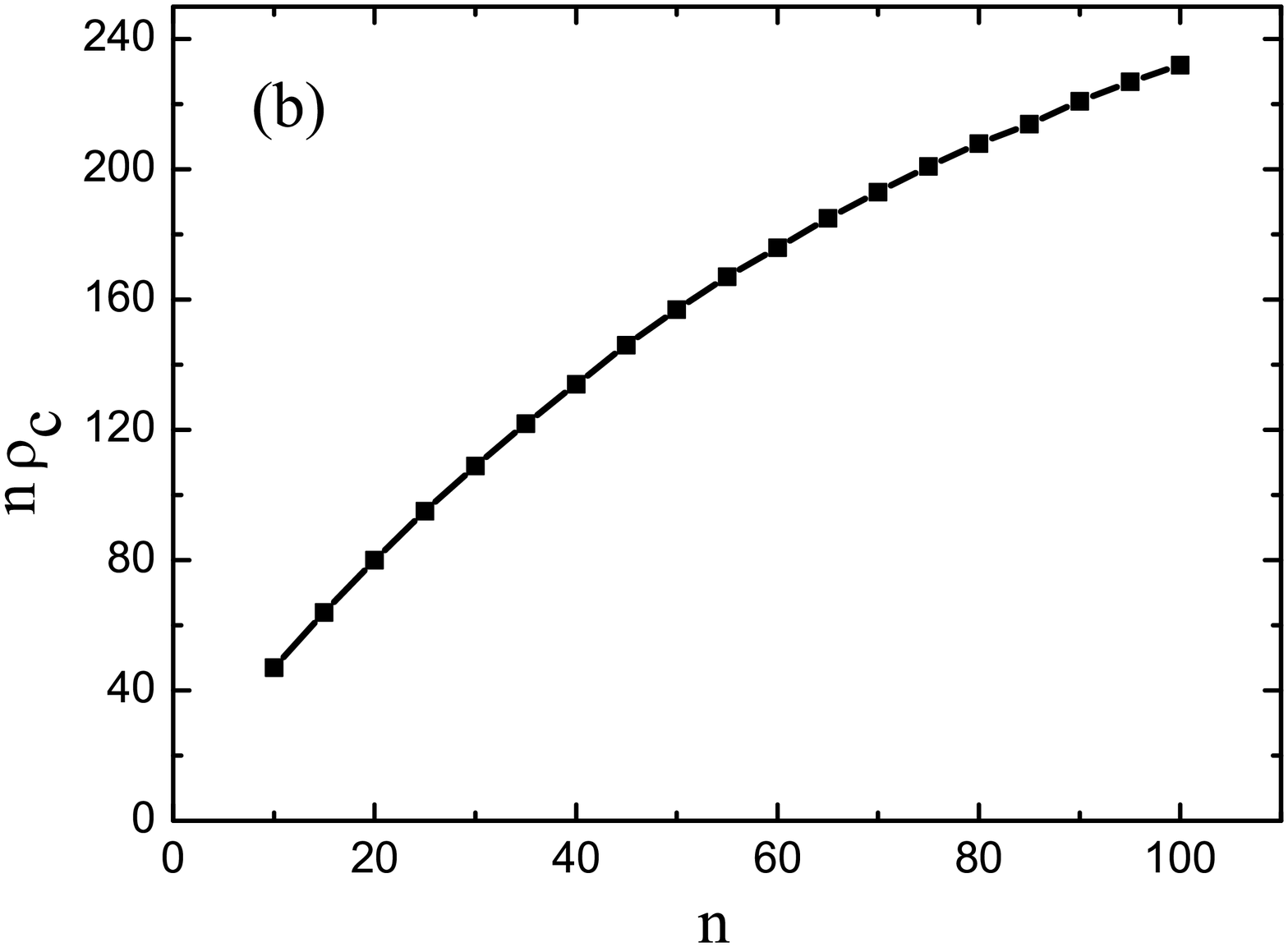}
\caption{(a) Critical packet generation rate $\rho_c$  and (b) total number of new packets per time step  $n\rho_c$ vs. number of users $n$. The backbone is optimized by the DBRS with $ P = 0.25$.
 The number of base stations is $ N = 1024$.   The delivery capacity of base station is $C =10$, and the user moving speed is $v=0.3$.   Each data point is the average of 1000 independent runs.}
\end{figure}

In Fig. 6, $n\rho_c$ represents the total number of inserted packets  when the traffic flow is at the critical point. With the increase of the number of users, the critical packet generation rate decreases (Fig. 6(a)).  When the number of users increases, more shortest paths  are used for packets transmission in the backbone, and this will increase the load of large-degree base stations, and thus decrease the critical packet generation rate. However, the  total number of  packets  the whole network can handle at  a time step increases  with the number of users (Fig. 6(b)).

Finally, we try to improve the network capacity by optimizing the routing protocol. Here we test the load-based shortest path protocol introduced in Section 3.1.  The results are shown in Fig. 7. The shape of the curves is the same as in Fig. 2. Also, we see that when there are multiple choices of next hops to a destination, it is better to choose the one with the minimum traffic load to deliver the packet, since the corresponding $\rho_c$ is larger than random choosing and that of  the maximum traffic load, as shown in Fig. 7.

\begin{figure}
\centering
\includegraphics[width=3in,height=2.25in]{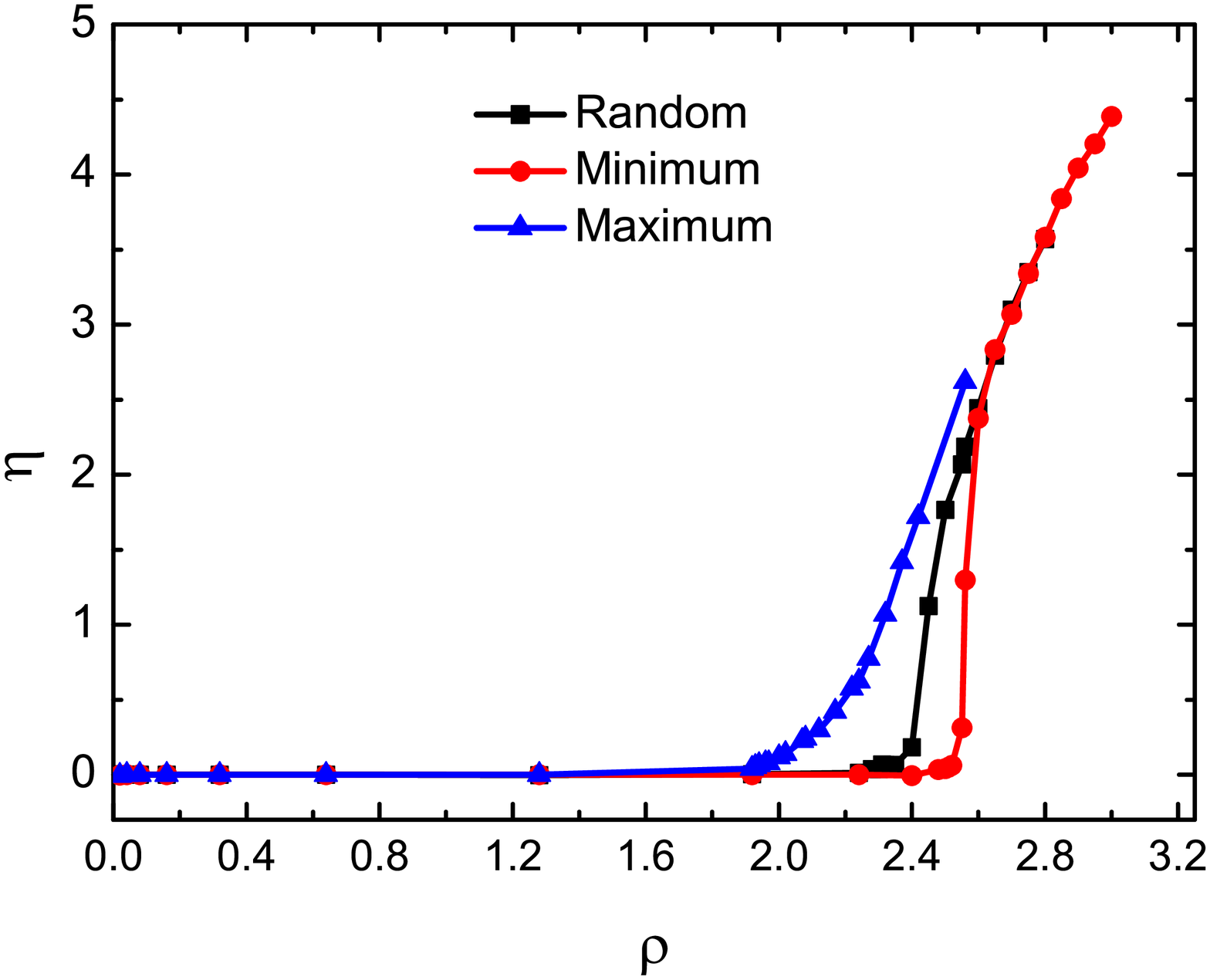}
\caption{Order parameter $\eta$ vs.  packet generation rate $\rho$ for three different routing strategies.
 The backbone is optimized by the DBRS with $ P = 0.25$. ``Random" means when there are multiple shortest paths between two nodes, we randomly pick one of them as the transmission path. ``Minimum" represents selecting the shortest path in which the next hop has the least  traffic load ( see Eq. (2)).  ``Maximum" represents selecting the shortest path in which the next hop has the most traffic load.
 The number of base stations is $ N = 1024$.   The delivery capacity of base station is $C =10$. The number of users is $n=100$, and the user moving speed is $v=0.3$.   Each data point is the average of 1000 independent runs.}
\end{figure}

\section{Conclusions}
Previously, researchers only discussed the network capacity of fixed or dynamic networks. In this paper, we study the network capacity of hybrid network composed of both fixed nodes and mobile nodes, a prototype of mobile communication networks and many others. In our model, the hybrid network includes a backbone consisting of fixed base stations for transmitting the packets. The mobile users generate new packets and remove arrival packets. We observe the phase transition of the traffic flow from free flow state to traffic congestion state. We obtain the optimal moving speed of users corresponding to the maximum network capacity. We propose a direction-based rewiring strategy to optimize the topological structure of the backbone to improve the network capacity. We also optimize the shortest path protocol by considering the traffic load of base stations. Note that besides information transmission, there are many other interesting problems of hybrid networks needed to be explored, such as viruses spreading, synchronization, etc.

\section*{Acknowledgments}
This work was  supported by the National Natural Science Foundation of China (Grant No. 61304154), the Specialized Research Fund for the Doctoral Program of Higher Education of China  (Grant No. 20133219120032).




\section*{Appendix A. The code for constructing the  lattice}
The original lattice is constructed with the following code:
\begin{lstlisting}[ language=C++]
function build_lattice_graph(edge) {
  let point(x, y) = x * edge + y;
      // total |nodes| = edge * edge
  graph = {};
  for (i = 0; i < edge; ++i)
    for (j = 0; j < edge; ++j) {
      if (i + 1 < edge)
        graph.add_edge(point(i, j), point(i+1, j));
      if (j + 1 < edge)
        graph.add_edge(point(i, j), point(i, j+1));
    }
  return graph;
}
\end{lstlisting}

\section*{Appendix B. The code for the DBRS}
In the DBRS, every base station in the original lattice rewires its right link and down link respectively with probability $P$. The new right (down) base station of the right (down) link is randomly chosen from the right (down) direction. The code for  the DBRS is given as follows:
\begin{lstlisting}[ language=C++]
function DBRS_optimization(edge, possibility) {
  let point(x, y) = x * edge + y;
  graph = {};
  for (i = 0; i < edge; ++i)
    for (j = 0; j < edge; ++j) {
      v1 = select any one point(x, y) where x > i && y > i
      if (v1.exists() && random(0, 1) < possibility)
        graph.add_edge(point(i, j), v1);
      else if (i + 1 < edge)
        graph.add_edge(point(i, j), point(i+1, j));

      v2 = select any one point(x, y) where x > i && y > i
      if (v2.exists() && random(0, 1) < possibility)
        graph.add_edge(point(i, j), v2);
      else if (j + 1 < edge)
        graph.add_edge(point(i, j), point(i, j+1));
    }
  return graph;
}
\end{lstlisting}

\section*{Appendix C. The code for calculating $\rho_c$}
We use the bisection method to obtain the critical packet generation  rate $\rho_c$, the code of which is as follows:
\begin{lstlisting}[ language=C++]
function get_rho_c {
  threshold = 1.0;
  rho_l = 1, rho_h = 2, rho_mid = 1;

  // O(logN) to get the upper bound of rho_c
  while (get_increasing_speed(rho_h) < threshold)
    rho_h = rho_h + rho_h;

  // O(logN) to get the exact value of rho_c
  while (rho_l < rho_h) {
    rho_mid = (rho_l + rho_h) / 2;
    if (get_increasing_speed(rho_mid) < threshold)
      rho_l = rho_mid + 1;
    else
      rho_r = rho_mid;
  }
  // we are sure that rho_l and rho_h are equal
  return rho_l;
}
\end{lstlisting}


\end{document}